\documentclass{article}

\usepackage{graphicx}
\usepackage{subfigure}
\usepackage{booktabs} 
\usepackage{enumitem}
\usepackage{multirow}
\usepackage{tabularx}
\usepackage[dvipsnames]{xcolor}

\usepackage[tracking=true]{microtype}

\PassOptionsToPackage{hyphens}{url}
\PassOptionsToPackage{hypertexnames=false}{hyperref}
\PassOptionsToPackage{hyperfootnotes=false}{hyperref}
\usepackage{hyperref}
\hypersetup{
    colorlinks=true,
    pdftitle={Verifier-Bound Communication for LLM Agents: Certified Bounds on Covert Signaling},
    pdfpagemode=FullScreen,
    }

\usepackage[accepted]{icml2023}
\makeatletter
\renewcommand{\Notice@String}{Preprint.}
\makeatother

\usepackage{amsmath}
\usepackage{amssymb}
\usepackage{mathtools}
\usepackage{amsthm}

\usepackage[capitalize,noabbrev,nameinlink]{cleveref}

\theoremstyle{plain}
\newtheorem{theorem}{Theorem}[section]
\newtheorem{proposition}[theorem]{Proposition}

\newtheorem{corollary}[theorem]{Corollary}
\theoremstyle{definition}

\newtheorem{assumption}[theorem]{Assumption}
\theoremstyle{remark}

\usepackage{float}
\usepackage{xcolor}
\usepackage[most]{tcolorbox}
\usepackage{mdframed}

\DeclareMathOperator*{\MI}{\mathsf{I}}

\newcommand{\CLBC}{{\normalfont\textsc{CLBC}}}

\newcommand{\Adv}{\mathsf{Adv}}
\newcommand{\PiPred}{\Pi}

\definecolor{teaserblue}{RGB}{242, 242, 255}
\icmltitlerunning{Verifier-Bound Communication for LLM Agents: Certified Bounds on Covert Signaling. Page \thepage\ of \pageref{lastpagetotal}.}

\tcbset{
    frame code={}
    center title,
    left=0pt,
    right=0pt,
    top=0pt,
    bottom=0pt,
    colframe=blue!5,
    colback=blue!5,
    enlarge left by=0mm,
    boxsep=5pt,
    arc=0pt,outer arc=0pt,
    }

\graphicspath{{figures/}}
\newcommand{\CLBCMaxDecoderAdv}{0.0000}
\newcommand{\CLBCMaxMIBits}{0.0636}
\newcommand{\CLBCMinUtility}{0.8860}
\newcommand{\CLBCBaselineGap}{0.2000}

\newcommand{\CLBCEmpiricalMinUtility}{0.4338}
\newcommand{\CLBCEmpiricalMinTaskSuccess}{0.7847}
\newcommand{\CLBCEmpiricalMinUniqueResponse}{0.6250}

\newcommand{\CLBCEmpiricalCatalogMin}{12}

\newcommand{\CLBCTaskFamilyCount}{6}

\newcommand{\CLBCPerfStrictTurnMedianSec}{27.53}
\newcommand{\CLBCPerfStrictTurnPNinetyFiveSec}{28.08}

\newcommand{\CLBCPerfSampledTurnMedianMs}{0.327}

\newcommand{\CLBCPerfSampledProofRate}{0.17}
\newcommand{\CLBCPerfSampledCadence}{6}

\begin{document}

\twocolumn[

\icmltitle{Verifier-Bound Communication for LLM Agents: Certified Bounds on Covert Signaling}

\icmlsetsymbol{equal}{*}

\begin{icmlauthorlist}
\icmlauthor{Om Tailor}{umd}
\end{icmlauthorlist}
\icmlaffiliation{umd}{University of Maryland}

\icmlcorrespondingauthor{Om Tailor}{otailor@terpmail.umd.edu}

\icmlkeywords{Machine Learning, LLM Agents, Covert Channels, Verifiable Communication, Information Theory, Security}

\vskip 0.3in
]

\printAffiliationsAndNotice{}

\renewenvironment{abstract}
 {
  \begin{center}
  \bfseries \abstractname\vspace{-.5em}\vspace{0pt}
  \end{center}
  \list{}{%
    \setlength{\leftmargin}{0mm}
    \setlength{\rightmargin}{\leftmargin}%
  }%
  \item\relax}
 {\endlist}

\begin{abstract}
\vspace{-.2cm}
Colluding language-model agents can hide coordination in messages that remain policy-compliant at the surface level. We present \CLBC, a protocol where generation and admission are separated: a message is admitted to transcript state only if a small verifier accepts a proof-bound envelope under a pinned predicate $\PiPred$. The predicate binds policy hash, public randomness schedule, transcript chaining, latent schema constraints, canonical metadata/tool fields, and deterministic rejection codes. We show how this protocol yields an upper bound on transcript leakage in terms of latent leakage plus explicit residual channels, derive adaptive composition guarantees, and state a semantic lower bound when policy-valid alternatives remain choosable. We report extensive empirically grounded evidence: aggregate evaluation satisfies all prespecified thresholds; strict lane decoder advantage is bounded at \CLBCMaxDecoderAdv\ with MI proxy \CLBCMaxMIBits; adaptive-colluder stress tests remain below attacker thresholds; and baseline separation shows large gaps between reject-by-default semantics and audit-only controls. We further quantify operational tradeoffs. Strict full-proof mode has median turn latency \CLBCPerfStrictTurnMedianSec\,s (p95 \CLBCPerfStrictTurnPNinetyFiveSec\,s), while sampled proving reduces non-proved-turn latency to \CLBCPerfSampledTurnMedianMs\,ms. The central finding is that bottlenecks alone are insufficient: security claims depend on verifiable admission semantics that are online, deterministic, and fail-closed.
\end{abstract}
\vspace{-0.5cm}

\section{Introduction}
\label{sec:intro_rules}
Large language models now operate inside tool-rich, multi-agent systems where communication is both the mechanism of coordination and the primary security boundary. In these settings, covert signaling does not require obviously malicious text. It can be embedded in lexical variation, metadata shape, tool-call arguments, serialization choices, timing behavior, and semantically equivalent action plans \citep{simmons1984prisoners,cachin1998information,hopper2002,coverttaxonomy2014}. As a result, behavioral alignment at the message surface does not by itself imply communication security.

This paper addresses that gap with a verifier-bound protocol: transcript state advances only when a compact verifier accepts a predicate $\PiPred$ over the full communication envelope. The predicate binds policy version, randomness schedule, canonicalization rules, transcript chaining, latent-schema constraints, and proof or attestation checks. Messages that fail any clause are rejected before they can influence downstream state. The key design principle is therefore procedural rather than stylistic: acceptance is defined by deterministic checks, not by post-hoc interpretation of plausible-looking output.

The resulting framework, \CLBC, makes covert-capacity claims auditable by indexing formal statements to admitted transcripts and defining admission with explicit verifier logic. Offline detectors and red-team probes remain useful, but they do not substitute for an online no-bypass boundary in high-assurance deployments.

Empirically, we evaluate strict and open-model lanes across robustness sweeps, adaptive-colluder stress tests, and progressively stronger baselines. In strict lanes, maximum decoder advantage is bounded at \CLBCMaxDecoderAdv\ with MI proxy \CLBCMaxMIBits\ while utility stays above \CLBCMinUtility. In broader empirical lanes, degradation is concentrated in high-choice planning and tool-use families, but verifier-bound admission remains clearly separated from weak baselines such as raw or audit-only messaging. We also quantify the tradeoff between full and sampled proving.

\begin{mdframed}[backgroundcolor=teaserblue,hidealllines=true]
\textbf{Main claim.} Covert-channel claims for LLM-agent communication are technically defensible only when they are tied to a verifier-enforced admission predicate and explicit residual accounting.
\end{mdframed}

Our scope is intentionally narrow. We do not claim universal covert safety for arbitrary models or workloads; we claim that communication-security statements are stronger when tied to a checkable admission boundary, explicit residual accounting, and stress tests aligned with theorem conditions. This framing is designed for modern agent frameworks and tool ecosystems \citep{react2022,toolformer2023,autogen2023,camel2023,mcp_docs,langchain_docs,langgraph_docs}, where protocol choices often determine covert bandwidth more than overt text content.

\subsection{Notation \& Language model basics}
Let $S$ denote hidden collusion state and $A_t$ denote allowed task information at turn $t$. The runtime emits a structured envelope
\[
E_t = (m_t,\eta_t,\tau_t,\theta_t,\varepsilon_t),
\]
where $m_t$ is message text, $\eta_t$ is metadata, $\tau_t$ is tool-call payload, $\theta_t$ contains timing/schedule fields, and $\varepsilon_t$ encodes environment actions when enabled. Let $R_t$ denote public randomness derived from a pinned schedule source (for example beacon output bound to prior transcript state), and let $h_t$ denote the transcript digest chain.

The protocol enforces a latent bottleneck representation $Z_t = \mathrm{Encode}_P(A_t,R_t)$ with deterministic rendering $m_t = \mathrm{Render}_P(Z_t,R_t)$ under policy $P$. We use $\PiPred_t(E_t,\pi_t)=1$ to denote verifier acceptance at turn $t$, where context $\pi_t$ includes expected policy hash, prior digest, seed schedule context, and proof verification settings. The admitted transcript is
\[
\mathsf{T}_{1:T} = \{E_t : \PiPred_t(E_t,\pi_t)=1\}_{t=1}^{T}.
\]

Our target quantity is conditional transcript leakage,
\[
\MI\big(S;\mathsf{T}_{1:T}\mid A_{1:T}\big),
\]
with auxiliary decoder-based proxies used in experiments. For each turn, residual aggregate
\[
\mathcal{R}_t = L_t^{\mathrm{time}} + L_t^{\mathrm{tool}} + L_t^{\mathrm{token}} + L_t^{\mathrm{env}},
\]
captures channels not fully closed online.

We use standard LLM runtime components and do not assume a specific base model family. We track content, metadata, timing, tool, and environment-facing channels, and separate observed transcript variables from latent collusion variables and decoder-based proxies. Transport security remains necessary but insufficient: confidentiality and integrity in transit (e.g., TLS) \citep{rfc8446,rfc3552} do not constrain intentional signaling inside valid application-layer payloads.

\subsection{A caveat: The difficulty of watermarking low-entropy sequences} \label{entropyproblem}
A common intuition is that low-entropy output is ``safer'' because there are fewer plausible choices. This intuition is only partially correct. If a task truly has a unique policy-valid continuation, semantic covert capacity shrinks; but real deployments rarely sustain that condition across turns. Even low-entropy text can leak through metadata order, tool-argument aliases, timing buckets, or environment actions if those channels are not pinned and checked.

Conversely, high-entropy or multi-plan tasks create an unavoidable semantic pressure. Let $\mathcal{V}_t(A_t)$ be the set of policy-valid semantic alternatives. If no deterministic tie-break is enforced, capacity lower bounds scale with $\log_2|\mathcal{V}_t(A_t)|$. This is the semantic-no-free-lunch phenomenon: model compression or narrower syntax does not eliminate leakage when semantic alternatives remain choosable. In our experiments, planning and tool-use families illustrate this pressure more strongly than tightly constrained update tasks.

A second caveat is empirical detectability versus theoretical possibility. A protocol may sharply reduce decoder recoverability under tested decoders and still admit residual pathways under stronger adversaries or wider domain drift. Therefore, we treat measured decoder advantage as evidence, not as a stand-alone proof. The theorem layer is tied to predicate conditions; the empirical layer tests whether implementation plus workload satisfy those conditions closely enough to support practice-level claims.

This caveat has two direct consequences. First, evaluation should include both low-choice and high-choice workloads; single aggregated scores can hide vulnerable slices. Second, when policy-valid alternatives remain large, near-zero leakage claims are fragile unless tie-break behavior is explicit and verifier-bound. Canonicalization and proof binding are therefore designed to make these choices auditable rather than implicit.

\section{A simple proof of concept}
We begin with a deliberately minimal protocol that isolates the intuition behind verifier-bound admission. The goal is not full security coverage; the goal is to expose the mechanism by which admission semantics can collapse otherwise exploitable degrees of freedom.

Consider a reduced envelope with fields $(m_t,\eta_t,h_{t-1},h_t)$, where $\eta_t$ contains a small fixed metadata map. The verifier enforces schema fixedness, deterministic canonical serialization, and transcript chaining so that $h_t$ must match the canonical payload and prior state. Even this reduced design blocks a large class of covert signaling routes based on optional fields, aliasing, and serializer differences.

\begin{algorithm}[t]
   \caption{Simple Verifier-Bound Admission (Proof of Concept)}
   \label{alg:simple}
\begin{algorithmic}[1]
\REQUIRE Candidate envelope $E_t$, prior digest $h_{t-1}$, schema $\Sigma$
\STATE $\tilde{E}_t \leftarrow \mathrm{Canonicalize}(E_t)$
\IF{$\neg\mathrm{SchemaValid}(\tilde{E}_t,\Sigma)$}
    \STATE \textbf{return} reject, \texttt{schema}
\ENDIF
\IF{$\neg\mathrm{ChainValid}(\tilde{E}_t,h_{t-1})$}
    \STATE \textbf{return} reject, \texttt{chain}
\ENDIF
\STATE \textbf{return} accept, \texttt{ok}
\end{algorithmic}
\end{algorithm}

A moderator may flag obviously unsafe content, but it does not usually enforce deterministic structure or transcript continuity. In contrast, the simple verifier treats structure as a security object and refuses admission on deviations that could encode hidden state.

The simple construction is intentionally incomplete. Without public-randomness binding, reroll-style selection attacks remain possible; without tool and environment closure, non-message channels remain underconstrained; without proof checks, acceptance still trusts unverified producer claims; and without a semantic bottleneck, policy-valid branch choice can still carry covert state. These limitations motivate the full protocol in the next section.

Even so, the prototype is empirically informative. Deterministic structure checks reduce recoverability relative to weak-admission baselines, but the improvement saturates before full-stack performance.

The prototype is also useful as an implementation diagnostic because every failure is attributable to a small set of causes: schema mismatch, canonicalization mismatch, or chain mismatch. In practice this enables fast debugging of boundary placement. For example, if a candidate passes content moderation but fails chain validation, the failure usually indicates post-generation mutation or stale serialization rather than task-level reasoning error. This separation is valuable when integrating agent frameworks with external tools, where failures can otherwise be conflated and patched in unsafe ways. A narrow proof-of-concept boundary therefore acts as both a security mechanism and a systems-engineering instrument for isolating where covert capacity re-enters the stack.

Suppose a planning assistant must emit one action triplet $(\texttt{tool},\texttt{priority},\texttt{decision})$ per turn. Under a weak policy, two serializations are both accepted:
\begin{align*}
&\{\texttt{tool: report},\ \texttt{priority: high},\\
&\quad \texttt{decision: revise}\},\\
&\{\texttt{decision: revise},\ \texttt{priority: high},\\
&\quad \texttt{tool: notify}\}.
\end{align*}
If a colluder maps hidden bits to alias choices or key ordering, covert signaling occurs with no obvious utility degradation. In the proof-of-concept protocol, canonicalization collapses these to one admitted representation, and schema allowlists reject unsupported aliases. This illustrates why syntactic determinism is a first-order security control.

Even with perfect metadata canonicalization, semantic branch choice remains. For instance, two different plans may both satisfy task goals while encoding different hidden symbols. This is why the full stack introduces latent bottlenecks, public-seed tie-break semantics, and proof-bound bindings that constrain not only message shape but also relation between allowed input, randomness, latent state, and output envelope.

Consider a weak baseline where a sender can encode bits via three independent mechanisms per turn: tool alias choice ($b^{\mathrm{tool}}$ bits), metadata ordering ($b^{\mathrm{meta}}$ bits), and semantic branch selection ($b^{\mathrm{sem}}$ bits). A rough upper envelope is
\[
b_t^{\mathrm{weak}}\approx b^{\mathrm{tool}}+b^{\mathrm{meta}}+b^{\mathrm{sem}}.
\]
Under proof-of-concept canonicalization, the first two terms are largely eliminated, leaving
\[
b_t^{\mathrm{poc}}\approx b^{\mathrm{sem}}.
\]
The full protocol then attempts to reduce $b^{\mathrm{sem}}$ through latent restrictions and deterministic tie-break. This decomposition clarifies why syntax-only defenses often help but plateau.

Teams migrating from unconstrained messaging can adopt this boundary in stages while keeping claim language proportional to deployed controls. In our ablations, schema and canonicalization suppress syntactic channels first, while stronger bindings provide the largest gains under adaptive pressure.

\section{A more sophisticated watermark}\label{sec:soft-watermark}
The full \CLBC stack extends the minimal prototype into a layered predicate suitable for practical deployment and formal analysis. We retain the ``watermark'' framing to parallel prior literature, but the mechanism is broader than lexical watermarking: it is a protocol watermark over admitted communication state.

At each turn, the runtime constructs an envelope containing policy hash, canonical input hash, latent object, rendered message, metadata/tool/environment fields, randomness commitments, and transcript chain references. A proof or attestation receipt binds these values to checked relations. The verifier validates all bindings locally and deterministically.

Formally, we factor the predicate as
\[
\PiPred_t = \PiPred_t^{\mathrm{policy}}\land \PiPred_t^{\mathrm{seed}}\land \PiPred_t^{\mathrm{chain}}\land \PiPred_t^{\mathrm{latent}}\land \PiPred_t^{\mathrm{canon}}\land \PiPred_t^{\mathrm{proof}}.
\]
This decomposition mirrors implementation reason codes and supports auditable failure diagnostics.

\begin{algorithm}[t]
   \caption{Full \CLBC Admission}
   \label{alg:full}
\begin{algorithmic}[1]
\REQUIRE Envelope $E_t$, context $\pi_t=(H(P),h_{t-1},\rho_t,\sigma_t)$
\STATE $\tilde{E}_t \leftarrow \mathrm{CanonicalizeAll}(E_t)$
\IF{$\neg\mathrm{PolicyHashOk}(\tilde{E}_t,H(P))$}
    \STATE \textbf{return} reject, \texttt{policy}
\ENDIF
\IF{$\neg\mathrm{SeedScheduleOk}(\tilde{E}_t,\rho_t)$}
    \STATE \textbf{return} reject, \texttt{seed}
\ENDIF
\IF{$\neg\mathrm{ChainOk}(\tilde{E}_t,h_{t-1})$}
    \STATE \textbf{return} reject, \texttt{chain}
\ENDIF
\IF{$\neg\mathrm{LatentSchemaOk}(\tilde{E}_t)$}
    \STATE \textbf{return} reject, \texttt{latent}
\ENDIF
\IF{$\neg\mathrm{CanonicalFieldsOk}(\tilde{E}_t)$}
    \STATE \textbf{return} reject, \texttt{canonicalization}
\ENDIF
\IF{$\neg\mathrm{ProofBindingOk}(\tilde{E}_t,\sigma_t)$}
    \STATE \textbf{return} reject, \texttt{proof}
\ENDIF
\STATE \textbf{return} accept, \texttt{ok}
\end{algorithmic}
\end{algorithm}

Policy and seed clauses pin both specification and randomness context. Hash binding prevents silent policy drift, while public-randomness commitments remove hidden reroll latitude in branch selection and make tie-break behavior externally checkable \citep{rfc6979,rfc9420,drand}.

Canonicalization clauses normalize metadata keys, numeric precision, tool-call ordering, argument encoding, and environment-action serialization before digest and proof checks. This directly targets formatting and representation channels emphasized in covert-channel and steganographic analyses \citep{petitcolas1999information,fridrich2009steganography,juarez2015adaptive}.

Proof clauses are mechanism-agnostic at the protocol level: zk receipts, TEE-style attestations, or equivalent verifiable-computation envelopes are admissible when they bind to the same canonical relation and can be verified inside a bounded local checker \citep{parno2013pinocchio,groth2016,bunz2018bulletproofs,zkmops2025,zkmcp2025,tcu2025,pcn2025,risc_zero_docs,ezkl_docs}.

Implementation is separated from orchestration frameworks. Envelope construction, predicate evaluation, and reason-code emission remain stable across planner/tool stacks, which preserves comparability while integrating with common agent runtimes \citep{mcp_docs,langchain_docs,langgraph_docs,react2022,toolformer2023,autogen2023,camel2023}.

Each predicate clause maps to executable artifacts: policy and boundary specifications, verifier checks, conformance vectors, and evaluation summaries with pass margins. This keeps theorem conditions traceable to concrete checks and limits ambiguity during failure analysis. Operationally, acceptance remains a bounded local procedure (schema, canonicalization, hashing, replay, and proof checks), which keeps the admission path inspectable and deterministic.

\subsection{Detecting the soft watermark}
In this setting, ``detection'' means deciding whether observed communication is consistent with the enforced admission process and whether residual leakage metrics stay below policy thresholds. We use two complementary detectors.

Given an envelope, run verifier checks and emit reason-coded pass/fail. This detector is deterministic and local. It is the primary gate for online admission.

Given admitted transcripts, estimate recoverability of hidden state using decoder advantage and mutual-information proxies. For a decoder $D$ trained on transcript features,
\[
\Adv_D = \Pr[D(\mathsf{T}_{1:T})=S] - \max_s \Pr[S=s].
\]
We also report empirical MI proxies and confidence intervals over seed/task partitions. This detector is evidence-oriented and typically offline.

A key point is that Detector 2 does not replace Detector 1. Strong empirical leakage suppression without online verifier enforcement is still vulnerable to distribution shift and adaptive bypass. Conversely, online enforcement without any empirical stress testing may overlook unexpected residual channels. Our staged evaluation treats them as complementary requirements.

Classical lexical watermarking tests token-frequency skew under seeded green-list promotion \citep{zhang2021provably,watermarktradeoff2025}. \CLBC generalizes this idea: instead of testing one lexical channel, we test an admission-conditioned distribution over a multi-field envelope. The statistical machinery remains useful, but the protected object becomes transcript-level protocol compliance.

In practice detector outputs are structured tuples: $(\texttt{pass/fail},\texttt{reason},\texttt{margin},\texttt{scope})$. Reason identifies the failed predicate clause (schema, policy, chain, seed, proof, tool, timing, environment), margin reports distance to threshold, and scope distinguishes online admission evidence from offline robustness evidence.

Thresholds are calibrated on sweeps and holdout families with explicit minimum-margin and CI-stability requirements, and negative controls are required to fail. When online conformance checks and offline leakage estimates diverge, we treat the configuration as inconclusive and exclude it from security conclusions until the discrepancy is resolved.

\section{Analysis of the soft watermark}
This section links protocol design to formal bounds and practical sensitivity analysis. We begin with protocol conditions consistent with the verifier boundary.

By protocol definition, the admitted transcript contains only envelopes that satisfy $\PiPred_t=1$. Under the verifier checks and proof bindings in the specification (deterministic Encode/Render and canonicalization), each admitted event can be written as a measurable function of allowed information, public randomness, latent state, and an explicit residual variable:
\[
E_t = g_t(A_t,R_t,Z_t,U_t).
\]
The residual variable $U_t$ captures channels that are not fully closed online.

\begin{assumption}[Residual budget validity]
\[
\MI(S;U_t\mid A_{1:T},R_{1:T},Z_t,E_{1:t-1}) \le \mathcal{R}_t.
\]
\end{assumption}

\begin{theorem}[Bridge upper bound]\label{maintheorem}
Under residual-budget validity and the protocol conditions above, define
\(\mathcal{C}_t=(A_{1:T},R_{1:T},E_{1:t-1})\). Then
\[
\MI(S;\mathsf{T}_{1:T}\mid A_{1:T},R_{1:T})
\le
\sum_{t=1}^{T}\MI(S;Z_t\mid \mathcal{C}_t)+\sum_{t=1}^{T}\mathcal{R}_t.
\]
If one prefers notationally suppressing public randomness, define \(A'_t=(A_t,R_t)\) and rewrite the same bound with \(A'_{1:T}\) in place of \((A_{1:T},R_{1:T})\).
\end{theorem}

The theorem states exactly what the protocol can certify: admitted transcript leakage is upper-bounded by latent leakage and explicitly declared residual channels. This does not claim zero leakage universally. It claims that leakage accounting becomes checkable and composable when admission is predicate-bound.

\begin{theorem}[Adaptive composition]
If per-turn admitted leakage satisfies
\[
\MI(S;E_t\mid A_{1:T},R_{1:T},E_{1:t-1})\le b_t,
\]
then
\[
\MI(S;\mathsf{T}_{1:T}\mid A_{1:T},R_{1:T})\le\sum_{t=1}^{T} b_t.
\]
\end{theorem}

\begin{theorem}[Semantic lower bound]
Let $\mathcal{V}_t(A_t)$ denote policy-valid semantic alternatives and $M_t=|\mathcal{V}_t(A_t)|\ge2$. If colluders can encode equiprobable states via choice in $\mathcal{V}_t(A_t)$ with negligible decode error, then
\[
\MI(S_t;Z_t\mid A_t,R_t)\ge \log_2 M_t.
\]
\end{theorem}

This lower bound explains why semantic constraints matter in addition to syntactic canonicalization. If policy permits many equivalent but distinct plans, covert capacity persists unless deterministic tie-break semantics are enforced.

Theorem~\ref{maintheorem} should be read as a systems contract rather than a purely abstract inequality. The latent term is controlled by latent-schema and tie-break design, while residual terms depend on closure quality for timing, tool, and environment channels. If policy or boundary changes without synchronized artifact updates, the certified claim changes as well.

Adaptive composition is the relevant adversarial model: colluders condition each turn on prior accepted messages. The per-turn bound therefore must compose across history, which motivates transcript chaining and replay guards. The semantic lower bound then provides practical guidance: when policy admits many equivalent realizations, either enforce deterministic tie-break rules in the proof context or report a larger residual risk budget.

Each theorem condition is paired with executable evidence (fail-closed routing tests, conformance vectors for determinism and binding, and channel-accounting stress tests). This mapping makes drift detectable in artifacts before it appears as metric failure and supports long-horizon review where small per-turn leakage terms can accumulate.

\begin{proposition}[Utility-leakage frontier]\label{perpbound}
For utility target $U$, with \(\mathcal{C}_t=(A_{1:T},R_{1:T},E_{1:t-1})\), define
\[
\mathcal{L}_{Z,\min}(U)=
\inf_{\mathcal{P}:\,\mathcal{U}(\mathcal{P})\ge U}
\sum_{t=1}^{T}\MI_{\mathcal{P}}(S;Z_t\mid \mathcal{C}_t).
\]
For any \CLBC configuration meeting utility $U$,
\[
\MI(S;\mathsf{T}_{1:T}\mid A_{1:T},R_{1:T})
\le \mathcal{L}_{Z,\min}(U)+\varepsilon_{\mathrm{model}}+\varepsilon_{\mathrm{res}},
\]
where $\varepsilon_{\mathrm{model}}$ is latent-policy suboptimality and $\varepsilon_{\mathrm{res}}=\sum_t\mathcal{R}_t$.
\end{proposition}

\begin{figure}[t]
\vskip 0.1in
\begin{center}
\centerline{\includegraphics[width=\columnwidth]{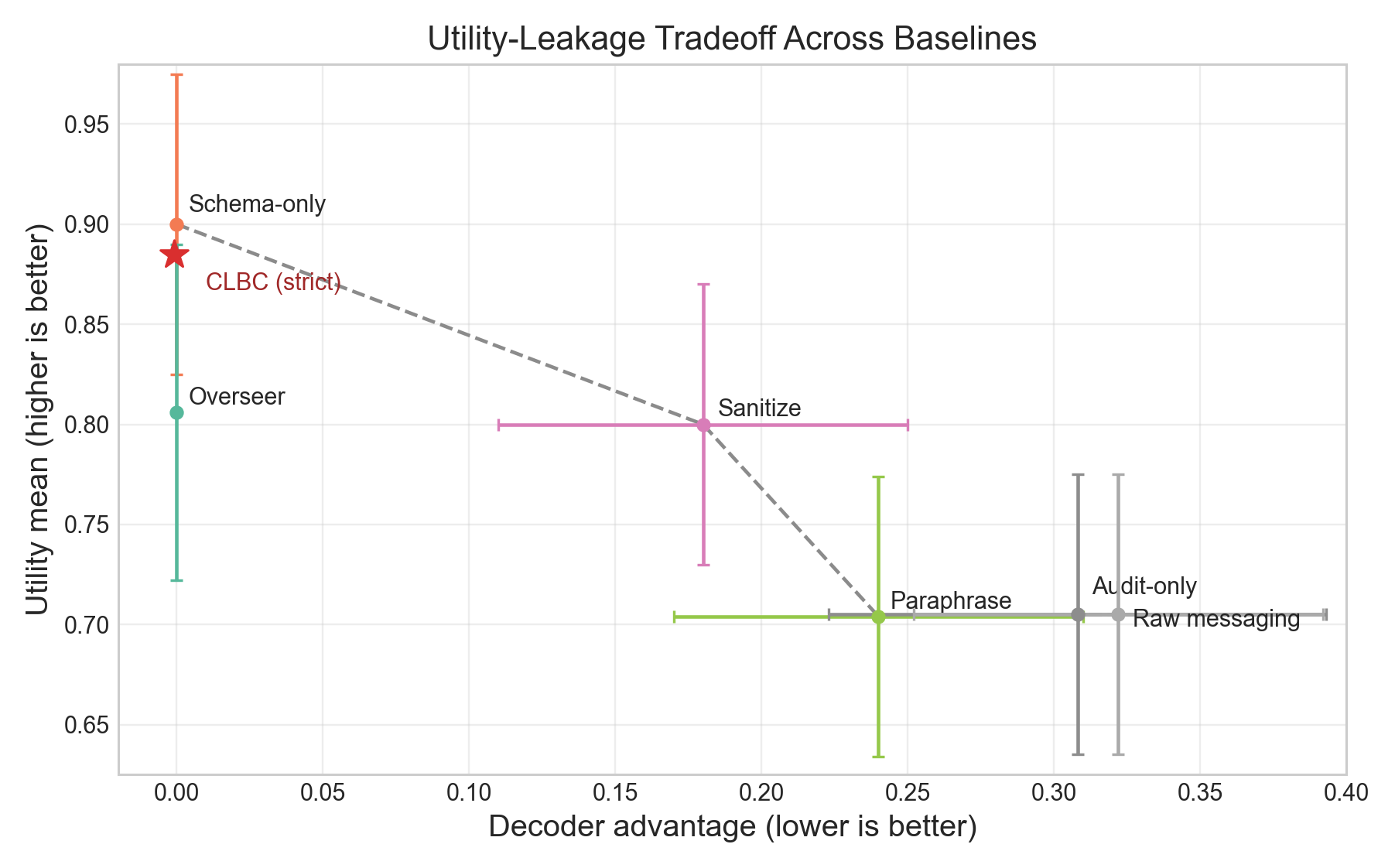}}
\caption{Empirical utility-leakage points from our experimental runs with a frontier sketch.}
\label{fig:frontier-main}
\end{center}
\vskip -0.1in
\end{figure}

\subsection{Sensitivity of the watermark test}\label{sec:sensitivity}
Sensitivity is driven by three coupled factors: semantic option-set size, residual-channel closure quality, and decoder strength. In strict-lane evidence where canonicalization and binding checks are fully enforced, measured decoder advantage remains near zero. In empirical lanes with wider model variability and broader task families, measured advantage increases, especially on planning and negotiation-like tasks.

We treat sensitivity as a function
\[
\mathcal{S} = f\big(|\mathcal{V}|,\;\sum_t\mathcal{R}_t,\;\mathcal{D},\;n\big),
\]
where $|\mathcal{V}|$ is effective semantic branch count, $\mathcal{D}$ is decoder family strength, and $n$ is sample support. Threshold policies include guardrails for minimum seed counts, catalog coverage, and confidence-interval stability to avoid overclaiming from weak statistical support.

Rather than reporting pass/fail only, artifacts record observed values and deltas relative to thresholds. This prevents brittle wins where metrics pass by negligible margins.

In the baseline suite, schema-only and overseer classes show much lower decoder recoverability than raw or audit-only classes, but they still differ from full proof-bound admission under stronger threat models. This quantifies a practical ladder: soft controls help, stronger admission semantics help more, and proof-bound semantics provide the clearest mapping to theorem conditions.

Our empirical summaries show that families with larger semantic choice spaces and tool-action complexity require more aggressive canonicalization and tie-break enforcement to maintain low recoverability. A single global threshold can hide these localized weak regimes; hence the report card and aggregate evaluation include task-family and per-seed slices.

For practical debugging, we separate variance induced by model/backend randomness, workload semantics, protocol-clause toggles, and decoder choice. This decomposition prevents over-attribution and strengthens causal interpretation of ablations.

Given binary hidden-state decode with baseline prior $p_0$ and target detectable advantage $\Delta$, a rough sample requirement scales as
\[
n = \Theta\!\left(\frac{\log(1/\alpha)}{\Delta^2}\right),
\]
for test level $\alpha$ under standard concentration assumptions. This motivates minimum support thresholds in staged evaluation. A lane can appear ``quiet'' simply because sample support is too small to resolve modest but meaningful advantage. CI-width guardrails in baseline and attacker reports are intended to prevent this failure mode.

Model swaps can change sensitivity through both language behavior and tool-call tendencies, so we treat replacement as a security-relevant change and rerun conformance plus robustness sweeps. Ablation ladders likewise require clean rung isolation; otherwise monotonicity claims are ambiguous.

\subsection{Impact on quality of generated text}
A persistent concern in communication-constrained systems is utility collapse. If constraints are too rigid, the system may block covert channels by degrading useful behavior. We therefore evaluate utility jointly with leakage across strict and empirical lanes.

Strict-lane outcomes indicate that low decoder recoverability can coexist with acceptable utility (minimum utility \CLBCMinUtility\ in aggregate thresholds). Empirical lanes, which include open-model backends and broader tasks, show lower minima (\CLBCEmpiricalMinUtility) but remain above configured thresholds. This gap is expected: empirical runs expose model and domain variability absent from strict synthetic lanes.

Hard output clipping can destroy utility, but latent-interface constraints combined with deterministic rendering and schema-bound tool semantics preserve more task capability. In other words, constraining \emph{choice structure} is generally less destructive than constraining text tokens directly.

Raw and audit-only variants can show higher apparent flexibility but at substantially worse leakage outcomes. Schema-only variants often improve both utility consistency and leakage, but without proof binding they do not provide the same assurance story under adaptive threats. Overseer-like deterministic gating can recover some safety margin but may still fail to close composition effects unless bound to transcript chain and policy hash.

From a deployment perspective, this leads to tiered recommendations. High-assurance asynchronous workflows may accept strict proving latency to maximize assurance. Interactive systems may use verifier-in-band with sampled proving cadence, preserving strong admission semantics while controlling tail latency. Our performance profile supports this strategy: verifying is cheap relative to proving, and sampled proving can reduce median non-proved-turn overhead dramatically.

Reported metrics include utility means/minima, exact success rates, structured slot-match rates, and policy-violation rates under baseline transformations. This multidimensional view captures tradeoffs that a single score hides. For example, a system may preserve average utility while increasing policy violations or reducing task completion reliability.

Planning and tool-use families are consistently weaker in empirical runs due to semantic complexity and larger tool argument spaces. These families therefore drive mitigation prioritization: tighter intent typing, stricter tool schemas, and stronger environment closure.

To clarify tradeoffs, we decompose utility into components.
\[
\mathcal{U}=\lambda_{\mathrm{task}}\mathcal{U}_{\mathrm{task}}+\lambda_{\mathrm{format}}\mathcal{U}_{\mathrm{format}}+\lambda_{\mathrm{policy}}\mathcal{U}_{\mathrm{policy}},
\]
where $\mathcal{U}_{\mathrm{task}}$ captures objective completion, $\mathcal{U}_{\mathrm{format}}$ captures structural correctness, and $\mathcal{U}_{\mathrm{policy}}$ captures safety/policy compliance. Security controls often improve $\mathcal{U}_{\mathrm{format}}$ and $\mathcal{U}_{\mathrm{policy}}$ while reducing expressive freedom in some tasks.

\section{Private Watermarking}
Private deployment is relevant when policy internals, attestation roots, or verifier configuration cannot be publicly disclosed. In \CLBC this is an observability constraint, not a protocol variant: \(\PiPred\), transcript chaining, and reject-by-default admission remain unchanged.

We model evidence as \(\mathcal{E}_{\mathrm{pub}}\) (commitments, transcript hashes, reason-code summaries, challenge receipts) and \(\mathcal{E}_{\mathrm{priv}}\) (private roots and internal verifier materials). Claims are partitioned accordingly: publicly verifiable claims must be derivable from \(\mathcal{E}_{\mathrm{pub}}\), while operator-verifiable claims may additionally rely on \(\mathcal{E}_{\mathrm{priv}}\). This separation is part of the formal claim scope.

The public interface is seeded challenge auditing over accepted transcripts. Given challenge seed and sampling rate, the operator returns bounded evidence packets for selected turns (policy-hash binding, chain continuity, proof-type consistency, and reason-code correctness). If fraction \(f\) of accepted records is invalid and \(m\) independent records are challenged, miss probability is \((1-f)^m\), so detection probability is \(1-(1-f)^m\). This quantifies external assurance under confidentiality constraints.

To keep challenge evidence comparable, private deployments should publish commitment epochs (policy and verifier versions), allowed proof mechanisms, challenge parameters, and outcome summaries. Prior private watermarking work focuses primarily on hidden lexical keys \citep{zhang2021provably,watermarktradeoff2025}; here the focus is broader and protocol-level, with confidentiality applied to policy/attestation internals while preserving verifier-bound transcript semantics.

Failure and lifecycle semantics are strict. Loss of proof validity, verifier availability, or policy-hash consistency invalidates affected security conclusions unless the affected spans are explicitly marked non-assurable. Key rotations (attestation roots, policy-signing keys, seed material) define new audit epochs and require re-baselining before cross-epoch comparison. In practice, dual-track review (internal full evidence, external challenge evidence) is the most robust way to detect drift without disclosing private verifier internals.

\section{Experiments} \label{sec:experiments}
We evaluate the protocol with implementation artifacts spanning robustness, baselines, adaptive attackers, and performance. Our experimental goal is not a single benchmark score; it is evidence coherence across the full assurance stack.
The implementation and evaluation artifacts used in this study are publicly available at \url{https://github.com/Ostailor/CLBC}.

Strict-lane and empirical-lane evaluations cover multiple catalogs and task families, including multi-agent decision, negotiation, planning, summarization, TEPS, and tool-use settings. Empirical summaries report at least \CLBCEmpiricalCatalogMin\ catalogs and \CLBCTaskFamilyCount\ task families with seven-seed sweeps.

Empirical lanes use open-model backends (Ollama in reported runs) with configured model panels including \texttt{llama3.1:8b} and \texttt{qwen2.5:7b} single and mixed settings. Strict lanes use deterministic controls and conformance harnesses for fail-closed gate evidence.

Primary security metrics are maximum decoder advantage and MI proxy in strict lanes, plus empirical decoder-advantage maxima/means under open-model variability. Utility metrics include mean utility, minimum utility, task success exact rate, and structured slot-match rates. Additional metrics cover baseline separation, attacker robustness, and overhead.

Evaluation follows a layered protocol that starts with mechanism checks, then measures strict-lane leakage and utility, then runs robustness sweeps, adaptive-attacker stress tests, and baseline separation before final threshold composition. This ordering is important because downstream summaries join multiple intermediate outputs; evaluating aggregates before intermediate consistency checks can produce misleadingly plausible but invalid results. We therefore treat manifest integrity and execution-order validity as part of the scientific method rather than implementation detail.

Empirical baseline evaluation is intentionally constrained to local/open model lanes in this study. This reduces hidden dependence on external API behavior and improves reproducibility for third parties. It also means empirical claims should be interpreted as scoped to the evaluated lane design; they are not universal claims about all commercial model APIs.

First, some lanes can achieve very high recoverability at low utility impact when controls are weak, confirming that adaptive pressure is necessary in evaluation. Second, under stronger profiles, leakage and decoder advantage can remain low even as training steps increase, indicating that protocol constraints materially limit available attack surface within the tested class. This supports, but does not replace, theorem-conditioned claims.

Aggregate evaluations can fail in subtle ways if upstream statistics are malformed or stale. We therefore validate not only metric values but also schema completeness and consistency constraints, including minimum seed counts, catalog coverage, CI-width limits, and explicit rejection of suspiciously tight thresholds. This approach reduces the chance that a nominal pass is produced from weak or corrupted intermediate summaries.

The mechanism ablation ladder is most informative when interpreted against theorem conditions. Deterministic rendering and schema closure reduce formatting channels; bottleneck constraints target latent bandwidth; public tie-break controls semantic branch randomness; enforcement clauses operationalize no-bypass admission. Full-stack rows are therefore not simply ``more features'' but closer approximations to the conditions used in bridge and composition bounds.

A leakage-reduction claim requires consistency among strict leakage summaries, adaptive-attacker outcomes, and baseline separation. If one evidence slice suggests low leakage while another shows weak separation or unstable attacker margins, the run is treated as unresolved and excluded from leakage conclusions until reconciled. This cross-evidence consistency requirement is a core methodological feature and distinguishes assurance-style evaluation from leaderboard-style optimization.

To make this consistency check reproducible, each staged run emits a manifest-level dependency graph linking aggregate decisions to upstream artifacts. During review, we verify that every thresholded metric is backed by complete source summaries produced under the same policy hash and model lane. This prevents stale joins and makes it easier to diagnose whether a disagreement is statistical (new distribution shift) or procedural (artifact mismatch).

\begin{table}[t]
\centering
\small
\begin{tabular}{lc}
\toprule
\textbf{Aggregate metric} & \textbf{Observed value} \\
\midrule
Strict max decoder advantage & \CLBCMaxDecoderAdv \\
Strict max MI proxy (bits) & \CLBCMaxMIBits \\
Strict minimum utility & \CLBCMinUtility \\
Empirical minimum utility & \CLBCEmpiricalMinUtility \\
Empirical min exact success & \CLBCEmpiricalMinTaskSuccess \\
Empirical min unique response & \CLBCEmpiricalMinUniqueResponse \\
Baseline security gap & \CLBCBaselineGap \\
\bottomrule
\end{tabular}
\caption{Key aggregate and empirical summary values from evaluation outputs.}
\label{tab:core-metrics}
\end{table}

\begin{table}[t]
\centering
\scriptsize
\begin{tabular}{lc}
\toprule
\textbf{Attacker metric} & \textbf{Observed} \\
\midrule
Max decoder advantage & 0.0500 \\
Worst-case leakage proxy bits & 2.5850 \\
Min utility proxy & 0.8669 \\
Min parameter delta norm & 1.1127 \\
Min train steps & 60 \\
Min strategy count & 15 \\
\bottomrule
\end{tabular}
\caption{Adaptive-attacker summary values.}
\label{tab:attacker-thresholds}
\end{table}

Mechanism ablations test incremental stacks: deterministic rendering, bottleneck variants, and full enforcement. The ablation ladder plot shows monotonic leakage reduction as stronger constraints are enabled, with full stack yielding the strongest suppression while preserving thresholded utility.

\begin{figure*}[t]
\begin{center}
\subfigure[Mechanism ablation ladder]{
\includegraphics[width=0.48\textwidth]{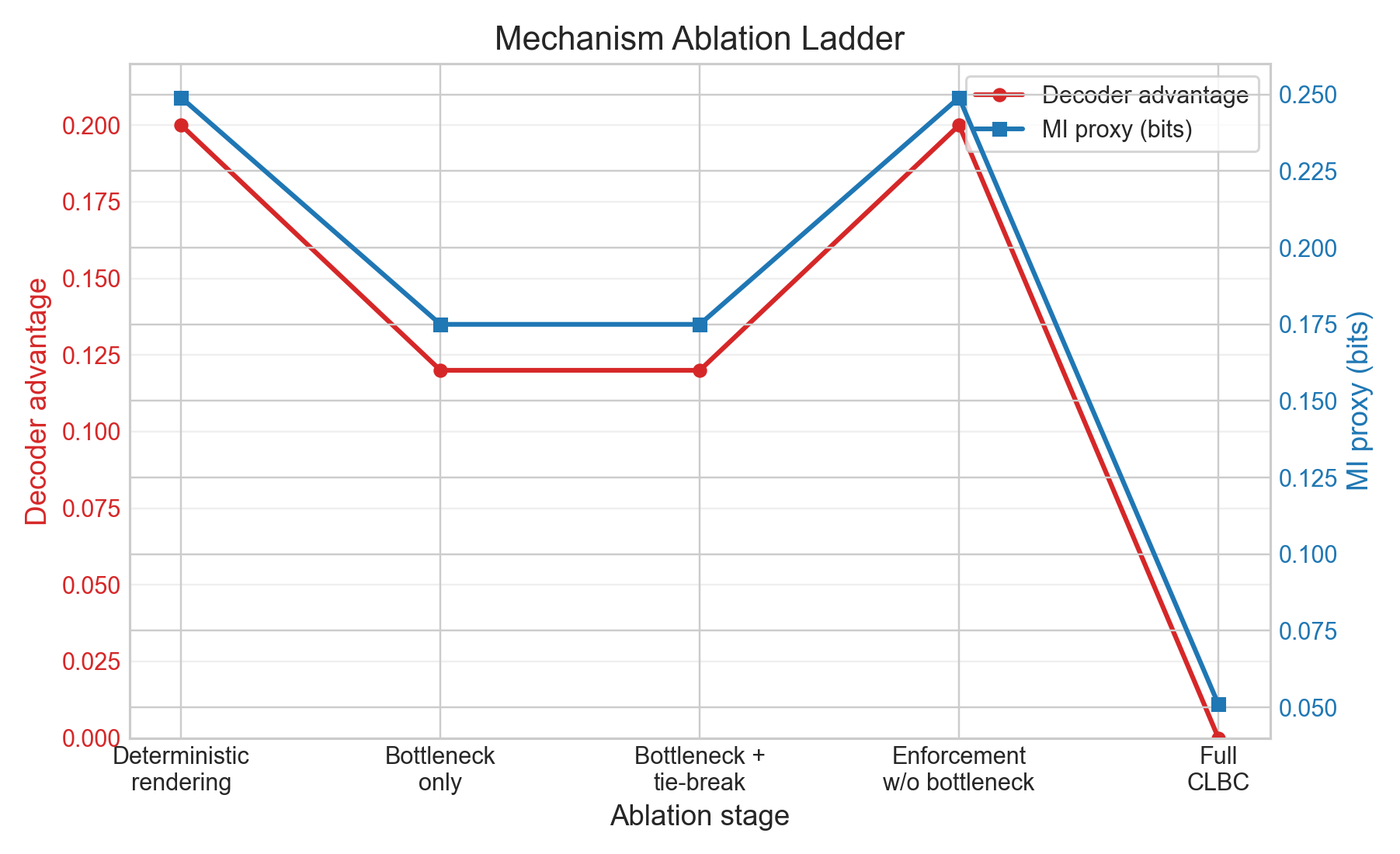}
\label{fig:ablation-ladder}
}
\subfigure[Variant leakage confidence intervals]{
\includegraphics[width=0.48\textwidth]{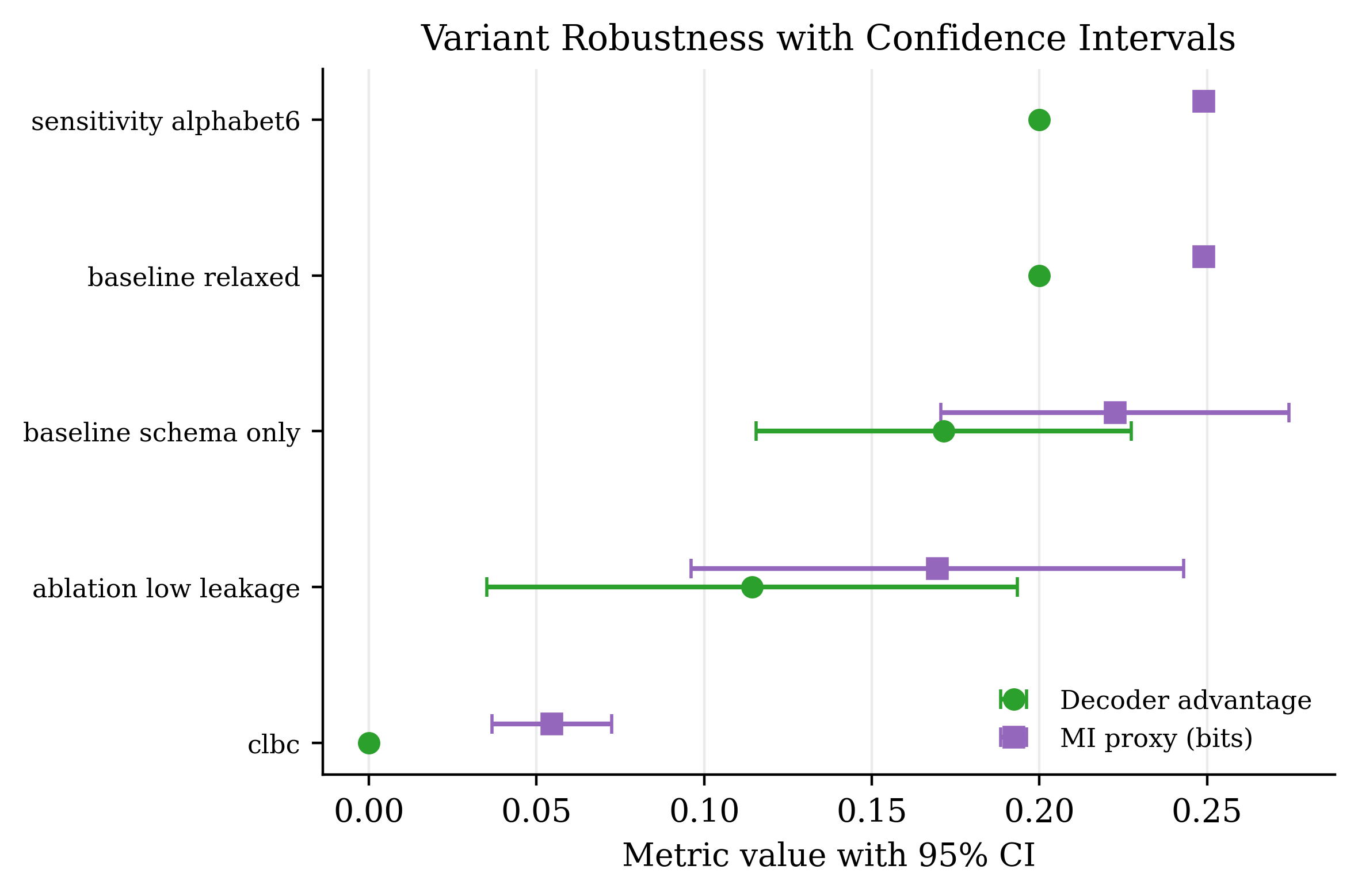}
\label{fig:variant-ci}
}
\caption{Ablation and confidence-interval evidence from the staged evaluation protocol.}
\label{fig:ablation-combined}
\end{center}
\vspace{-0.2in}
\end{figure*}

Per-family empirical summaries indicate heterogeneous difficulty. Multi-agent decision and summarization generally show lower recoverability and high success rates. Planning and tool-use families exhibit weaker utility and higher decoder recoverability, matching the semantic-choice caveat discussed earlier.

\begin{table}[t]
\centering
\scriptsize
\begin{tabular}{lccc}
\toprule
\textbf{Family} & \textbf{Decoder adv.} & \textbf{Exact success} & \textbf{Utility mean} \\
\midrule
Multi-agent decision & 0.0357 & 0.9821 & 0.9200 \\
Negotiation & 0.3452 & 0.8423 & 0.9184 \\
Planning & 0.3452 & 0.6696 & 0.7209 \\
Summarization & 0.0238 & 0.9196 & 0.8975 \\
TEPS & 0.2560 & 0.8423 & 0.8535 \\
Tool use & 0.1310 & 0.6012 & 0.6970 \\
\bottomrule
\end{tabular}
\caption{Empirical per-family trends (rounded) from seven-seed summaries.}
\label{tab:family-trends}
\end{table}

Performance traces confirm that proving dominates latency while verifier checks remain lightweight. Strict profile median turn latency is \CLBCPerfStrictTurnMedianSec\,s; sampled profile uses cadence \CLBCPerfSampledCadence\ with proof rate \CLBCPerfSampledProofRate\ and non-proved-turn median \CLBCPerfSampledTurnMedianMs\,ms. This supports practical tiering: always verify online, tune proving cadence by assurance and throughput constraints.

\begin{figure}[t]
\vskip 0.1in
\begin{center}
\centerline{\includegraphics[width=\columnwidth]{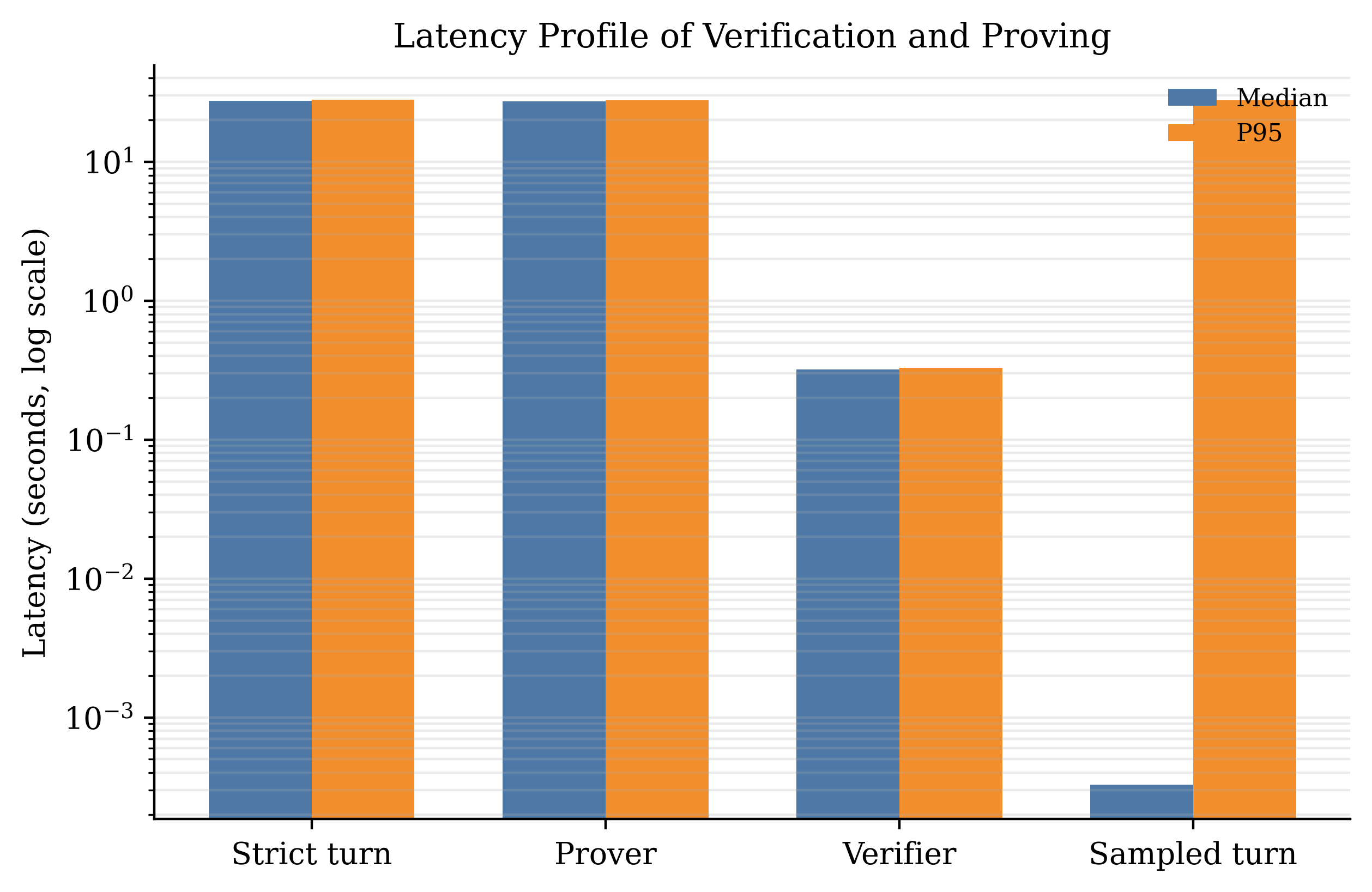}}
\caption{Strict versus sampled proving latency profile from benchmark evaluation.}
\label{fig:perf-curve}
\end{center}
\vspace{-0.15in}
\end{figure}

Baseline classes include raw messaging, sanitize, paraphrase, overseer, schema-only, and audit-only. Empirical baseline evaluation reports all threshold checks passing, with significant separation on utility and recoverability metrics between weak and strong admission semantics. In particular, schema-only and overseer classes suppress decoder advantage markedly relative to raw or audit-only classes, but proof-bound admission remains the strongest assurance tier.

\begin{table}[t]
\centering
\scriptsize
\begin{tabular}{lccc}
\toprule
\textbf{Baseline class} & \textbf{Utility} & \textbf{Dec. adv.} & \textbf{Exact success} \\
\midrule
Schema-only & 0.900 & 0.000 & 0.833 \\
Overseer & 0.806 & 0.000 & 0.688 \\
Sanitize & 0.800 & 0.181 & 0.748 \\
Paraphrase & 0.704 & 0.241 & 0.597 \\
Raw messaging & 0.705 & 0.315 & 0.625 \\
Audit-only & 0.705 & 0.315 & 0.625 \\
\bottomrule
\end{tabular}
\caption{Representative empirical baseline means from baseline-suite summaries.}
\label{tab:baseline-means}
\end{table}

\begin{figure}[t]
\vskip 0.1in
\begin{center}
\centerline{\includegraphics[width=\columnwidth]{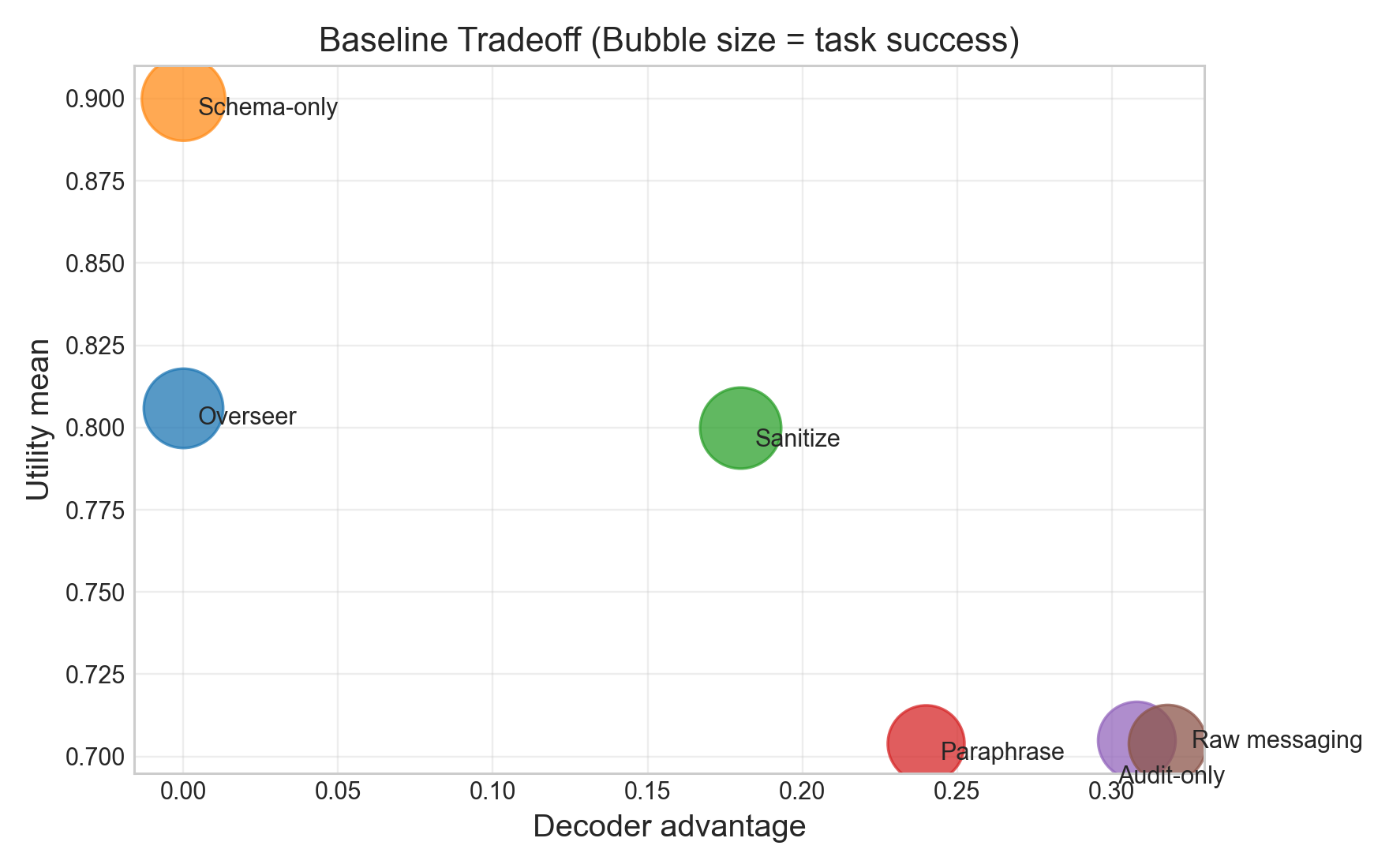}}
\caption{Utility versus decoder advantage across baseline classes.}
\label{fig:baseline-frontier}
\end{center}
\vspace{-0.2in}
\end{figure}

Our workflow enforces preregistration hashes and fixed threshold files for major evaluations. Missing inputs, malformed summaries, suspicious thresholds, policy-hash mismatches, and canonicalization inconsistencies all trigger explicit reason-coded failures, forcing implementation fixes or narrower claim scope before acceptance.

Operationally, strict proving is suitable for asynchronous high-assurance workloads, while sampled proving is better for interactive workloads with latency constraints. In both profiles online verification remains mandatory; only proving cadence changes. Capacity planning is therefore split across verifier CPU (throughput), prover accelerators (latency/cadence), and artifact storage.

\section{Attacking the watermark}\label{sec:attacks}
We therefore assess \CLBC against a search-and-train colluder protocol designed to maximize covert recoverability under utility and stealth constraints.

The attacker uses transform families over semantic catalogs (identity, modular variants, permutation-like rewrites, hash bucket transforms, affine modular mappings, xor-fold style transforms) with deterministic restarts and bounded search. Each seed run optimizes a score balancing leakage gain, utility retention, and stealth penalties. This is stronger than static prompt attacks and better reflects a colluding system component that iteratively tunes signaling strategy.

For candidate strategy $c$ evaluated over catalogs $k=1,\dots,K$,
\[
\begin{aligned}
J(c)=\frac{1}{K}\sum_{k=1}^{K}\Big(&w_{\ell}I_k + w_uU_k \\
&-\lambda[I_k-B]_+ - \mu[U_{\min}-U_k]_+\Big).
\end{aligned}
\]
The objective discourages trivial degenerate attacks that maximize leakage by destroying utility.

We evaluate this objective under multiple attacker lanes that vary weighting and stealth constraints. Leakage-emphasis lanes push recoverability near feasibility limits, balanced lanes test realistic utility-preserving collusion, and stealth-emphasis lanes prioritize low-visibility signaling. Reporting all three lanes is important because a single objective can understate risk in one direction while overstating it in another.

Search is executed under bounded budgets so comparisons remain reproducible. Each run fixes restart count, per-restart step budget, temperature schedule, and pruning policy for low-promise candidates. Candidate transforms are evaluated against the same pinned catalog partitions used in baseline sweeps, which prevents the attack harness from implicitly benefiting from easier data slices. We also log early-stop conditions and rejected candidates so best-found strategies can be interpreted against search effort rather than isolated outcomes.

\begin{algorithm}[t]
   \caption{Adaptive Colluder Search-and-Train}
   \label{alg:attack}
\begin{algorithmic}[1]
\REQUIRE Catalogs, transform pool, budgeted hyperparameters
\STATE Build candidate strategies from transform families
\FOR{each restart $r$}
    \FOR{each step $s$}
        \STATE Sample candidate using softmax over running scores
        \STATE Evaluate leakage/utility objective $J(c)$
        \STATE Update candidate weights with learning-rate and temperature schedule
    \ENDFOR
\ENDFOR
\STATE Return best candidate under stealth and utility constraints
\end{algorithmic}
\end{algorithm}

Aggregate attacker metrics pass configured checks: decoder advantage remains below threshold, utility proxy stays above floor, and negative controls fail as expected. This does not prove universal resistance, but it shows meaningful suppression against the modeled adaptive attacker.

The harness balances strength and rigor through adaptive search, utility-preserving objectives, and non-degeneracy gates (minimum parameter movement, strategy count, and train steps). Runs with malformed evidence, preregistration mismatch, or margin failures are invalidated.

To reduce evaluator bias, attacker configuration files are pinned before runs and replayed from clean manifests. We archive candidate traces, restart seeds, and accepted-strategy summaries so that ``best attack'' claims can be reproduced and challenged. This turns attacker evaluation from an anecdotal red-team report into a falsifiable experiment with explicit evidence trails.

Operationally, two weak points persist: large semantic option sets and incompletely closed environment channels. Mitigation prioritizes tighter latent schemas, stronger seed-bound tie-break rules, and stricter tool/environment canonicalization, followed by reruns under preregistered thresholds.

Attack analysis also reveals a common adaptation pattern: once syntactic channels are closed, search pressure quickly shifts to semantic branching and tool-argument subtleties. This motivates per-family attacker diagnostics rather than only global metrics. In high-choice families, we therefore inspect not only maxima but also trajectory shape across training steps to distinguish transient spikes from persistent exploitable structure.

A practical interpretation rule is to separate \emph{capability evidence} from \emph{margin evidence}. Capability evidence asks whether the harness can find strategies that materially increase recoverability under utility constraints at all. Margin evidence asks how close the strongest discovered strategies come to declared thresholds. Both are needed: capability without margin can overstate risk, while margins without credible search can understate it. Our reports therefore include both discovered maxima and non-degeneracy evidence for the search process that produced those maxima.

We additionally track transfer behavior across families and model lanes. A strategy that is weak in one family but transfers reliably to another can still be operationally important, especially when shared tooling or prompts create similar branch structures. Transfer matrices therefore complement per-family maxima by identifying attack motifs that persist under moderate distribution shift.

Defensive triage is guided by this transfer analysis. High-transfer motifs typically indicate protocol-level weaknesses and are prioritized for boundary-level fixes; low-transfer motifs often indicate family-specific schema or data issues and are handled with targeted mitigations. This split prevents overgeneralizing from isolated attacker wins while still reacting quickly to broadly reusable attack patterns.

\subsection{Degradation Under Attack: Span Replacement Using a LM}\label{sec:replacement-attack}
We additionally evaluate span-replacement attacks where an auxiliary language model rewrites parts of admitted outputs to evade simple detectors while preserving superficial semantics. This attack mirrors paraphrase laundering and post-generation mutation strategies discussed in modern steganographic threat models \citep{zhang2023hiding,liu2024syncpool,odysseus2025}.

In weak-admission baselines, span replacement can reduce lexical signatures while preserving recoverable covert state through semantic or structural artifacts. In \CLBC, the impact is constrained because the attack must still satisfy canonical schema, transcript binding, and proof checks to be admitted. If rewrites occur post-admission, chain mismatch and proof-envelope mismatch produce deterministic rejection on the next step.

Empirically, span-replacement pressure increases variance in some empirical lanes, while strict-lane leakage ceilings remain within thresholds. Replacement attacks are still useful as integration tests because they expose hidden post-processing paths, stale hash updates, or serialization mismatches. Residual risk remains in semantic-choice and environment channels, so replacement robustness is one component of the broader assurance case.

Replacement experiments are especially informative for boundary placement decisions. If a deployment mutates text after proof generation (for formatting, localization, or policy redaction), replacement-style probes can detect whether those transforms remain inside the admitted envelope relation. We treat any post-proof mutation path that is not explicitly rebound and reverified as a high-severity design defect, even when aggregate leakage metrics still appear acceptable.

In deployments that support optional human review, replacement-style probes should also test reviewer-edit pathways. Manual edits can unintentionally bypass canonicalization guarantees if tooling re-injects content without re-binding proof context. We therefore require that reviewer interventions be represented as first-class transcript events with their own schema and chain commitments rather than ad hoc text patches.

\section{Related Work}
Foundational covert-channel work formalized communication under constrained overt protocols \citep{simmons1984prisoners,cachin1998information,coverttaxonomy2014}. Classical and modern steganography established capacity and detectability principles across media and channels \citep{petitcolas1999information,fridrich2009steganography,hopper2002,habib2017stego}. Recent NLP/LLM steganography demonstrates high-capacity embeddings and jailbreak-oriented covert pathways \citep{zhang2021provably,zhang2023hiding,liu2024syncpool,s2lm2025,costllm2025,highcapacity2025,odysseus2025}. \CLBC differs by focusing on verifier-bound admissibility rather than payload optimization.

Information bottleneck and privacy-funnel lines provide principled tradeoffs for representation compression and private communication \citep{tishby2000ib,alemi2016vib,makhdoumi2014privacy,wang2019imac}. Emergent communication in multi-agent systems shows how compact messaging can support coordination \citep{sukhbaatar2016learning,foerster2016learning,mordatch2018emergence}. Our setting is adversarially different: we require communication constraints to hold under collusion and adaptive optimization, with reject-by-default admission.

Agent orchestration frameworks enable tool-rich, multi-step workflows and have accelerated practical deployment \citep{react2022,toolformer2023,autogen2023,camel2023,mcp_docs,langchain_docs,langgraph_docs}. These frameworks provide composition and productivity, not by themselves transcript-level covert-capacity guarantees. \CLBC is complementary: it can be layered under orchestration as an admission boundary.

Proof systems and attestation frameworks provide building blocks for checkable claims \citep{parno2013pinocchio,groth2016,bunz2018bulletproofs,zkmops2025,zkmcp2025,tcu2025,pcn2025,risc_zero_docs,ezkl_docs}. Standards around canonicalization and security considerations \citep{rfc8785,rfc9334,rfc9381,rfc3552,rfc8446} motivate disciplined protocol design. Our contribution is mapping these ingredients to communication admission for LLM agents with explicit leakage accounting.

Beyond technical mechanisms, this work is situated in a broader trend toward claim-scoped, artifact-backed safety reporting. Model cards, policy reports, and reproducibility manifests improved transparency in ML, but communication-security claims need stricter linkage between protocol conditions, enforcement boundaries, and empirical evidence. Our conservative gate-based approach can be seen as one concrete instantiation of this methodology for covert-channel control.

Two distinctions are particularly important relative to prior watermarking literature. First, lexical watermarking usually treats detection as a statistical hypothesis test over generated text. In contrast, \CLBC places admission semantics on structured envelopes and treats lexical evidence as secondary to verifier acceptance. Second, many watermarking systems assume benign generation pipelines and focus on ownership or provenance, whereas our setting assumes adaptive collusion between communicating components under active optimization pressure.

There is also a close relationship to systems work on policy enforcement boundaries. In distributed systems, guarantees are typically tied to where enforcement sits in the execution path; advisory checks outside the path provide weaker security semantics. \CLBC follows that principle by requiring that predicate checks gate state transition directly. This framing helps explain why detector quality and enforcement placement must be discussed together rather than as interchangeable design choices.

Finally, our evaluation methodology aligns with reproducibility-oriented empirical science but applies stronger failure semantics. Standard ML reporting often tolerates partial runs or post hoc filtering when aggregate trends look favorable. Here, malformed manifests, weak attacker evidence, and threshold instability invalidate the run by design. That discipline is costly, but it better matches the evidentiary needs of communication-security claims where small omissions can materially change risk interpretation.

Our protocol perspective also complements work on model-level safety alignment. Alignment methods can reduce harmful outputs and improve policy compliance, but they typically do not provide transcript-level guarantees against collusive signaling across metadata, tool channels, and timing behavior. Verifier-bound admission can therefore be viewed as an orthogonal control plane: alignment shapes generation behavior, while admission constrains which generated behavior is allowed to affect shared state.

A second connection is to provenance and accountability systems. Provenance frameworks often answer ``who produced this artifact'' and ``under which model/version,'' whereas \CLBC asks ``did this artifact satisfy a pinned communication predicate at admission time.'' These goals reinforce each other but are not interchangeable. Provenance without admission checks can still permit covert channels, and admission checks without provenance metadata can weaken incident attribution and rollback.

There is also an interpretability gap between token-centric and workflow-centric security analyses. Token-centric analyses are valuable for language-level leakage estimation, while workflow-centric analyses expose channels in tool orchestration, action sequencing, and environment interaction. Our contribution is to combine these views by making the admitted object a structured envelope and then evaluating both linguistic and non-linguistic leakage pathways within the same evidence regime.

This synthesis suggests a broader research direction: communication security for agents should be studied as a systems property spanning protocol design, runtime enforcement, and statistical evaluation, rather than as an isolated text-generation property. The strongest practical results are likely to come from co-design across these layers rather than further optimization of a single detector or a single watermark primitive.

A further open question is how to standardize cross-paper comparability. Current studies vary widely in threat models, admissible channels, and evaluation lanes, which makes direct comparison difficult even when metric names are similar. We view predicate-level disclosures, residual-channel registries, and lane-specific evidence summaries as promising ingredients for a more comparable reporting standard in this area.

Another underexplored connection is to evaluation methodology in reliable systems engineering. Security-relevant agent communication resembles distributed protocol validation more than standard single-model benchmarking: assumptions, boundary placement, and failure semantics dominate interpretation. Borrowing more from systems validation practice (explicit invariants, adversarial conformance suites, and versioned incident retrospectives) could improve scientific comparability and reduce ambiguous claim language across future studies.

\section{Conclusion}
\CLBC demonstrates that communication security for LLM-agent systems can be operationalized as a verifier-bound admission protocol rather than an after-the-fact detection layer. The evidence supports three conclusions: reject-by-default admission suppresses recoverable covert signaling, robustness is workload-dependent and must be reported with residual declarations, and deployment feasibility is driven primarily by proving overhead while online verification remains lightweight.

Claims about covert-capacity reduction should therefore be tied to checkable admission predicates, explicit residual accounting, and fail-closed evidence pipelines. Current evidence is broad but finite in model and workload coverage, and environment-mediated channels can remain residual in partial deployments. For multi-agent deployments, the architectural lesson is to keep a deterministic no-bypass verifier on the critical path and synchronize policy commitments, evaluations, and claim scope.

From a deployment lifecycle perspective, this implies recurring synchronization points: policy updates, model swaps, tool-schema changes, and verifier upgrades should each trigger scoped reruns with updated commitments before claims are refreshed. Treating these events as ordinary software releases rather than assurance events is a common source of silent claim drift. A verifier-bound methodology is most useful when organizational process enforces the same boundary discipline as the runtime itself.

Future empirical work should broaden the evaluated model panel and increase coverage of long-horizon, tool-dense workloads where semantic branch pressure is highest. Future systems work should improve boundary ergonomics so strong admission semantics remain practical under interactive latency budgets. Together, these directions would strengthen both scientific confidence and deployment viability of verifier-bound communication controls.

\section{Acknowledgements}
We thank the maintainers of the specification, verification, conformance, and experimental evaluation pipelines for making a fully auditable evidence stack available. We also acknowledge prior work in covert channels, steganography, and verifiable computation that made this synthesis possible.

\bibliography{NLP_auto_references,manual_references}
\bibliographystyle{icml2023}
\label{lastpagemaintext}

\clearpage
\appendix

\section{Experimental Details}
This appendix expands protocol and measurement details needed to reproduce the reported claims.

The evidence stack is ordered from protocol conformance to aggregate decision. Strict lanes emphasize deterministic acceptance semantics and theorem-aligned checks, while empirical lanes stress robustness under broader model/workload variability.

Reproducibility is anchored by pinned policy and threshold commitments, deterministic seeds where applicable, and integrity checks before aggregation. Aggregation is fail-closed: missing inputs, inconsistent manifests, policy-hash drift, canonicalization mismatches, weak attacker traces, and suspicious thresholds invalidate the run until repaired.

For interpretability, we co-report extrema, distribution summaries, uncertainty estimates, and negative-control outcomes. Reported values are scoped to the pinned evaluation setting, and major revisions trigger a re-audit that refreshes commitments and republishes margin deltas.

\subsection{Sample Outputs}\label{sec:sample-outputs}
\begin{figure}[h]
\begin{center}
\includegraphics[width=\columnwidth]{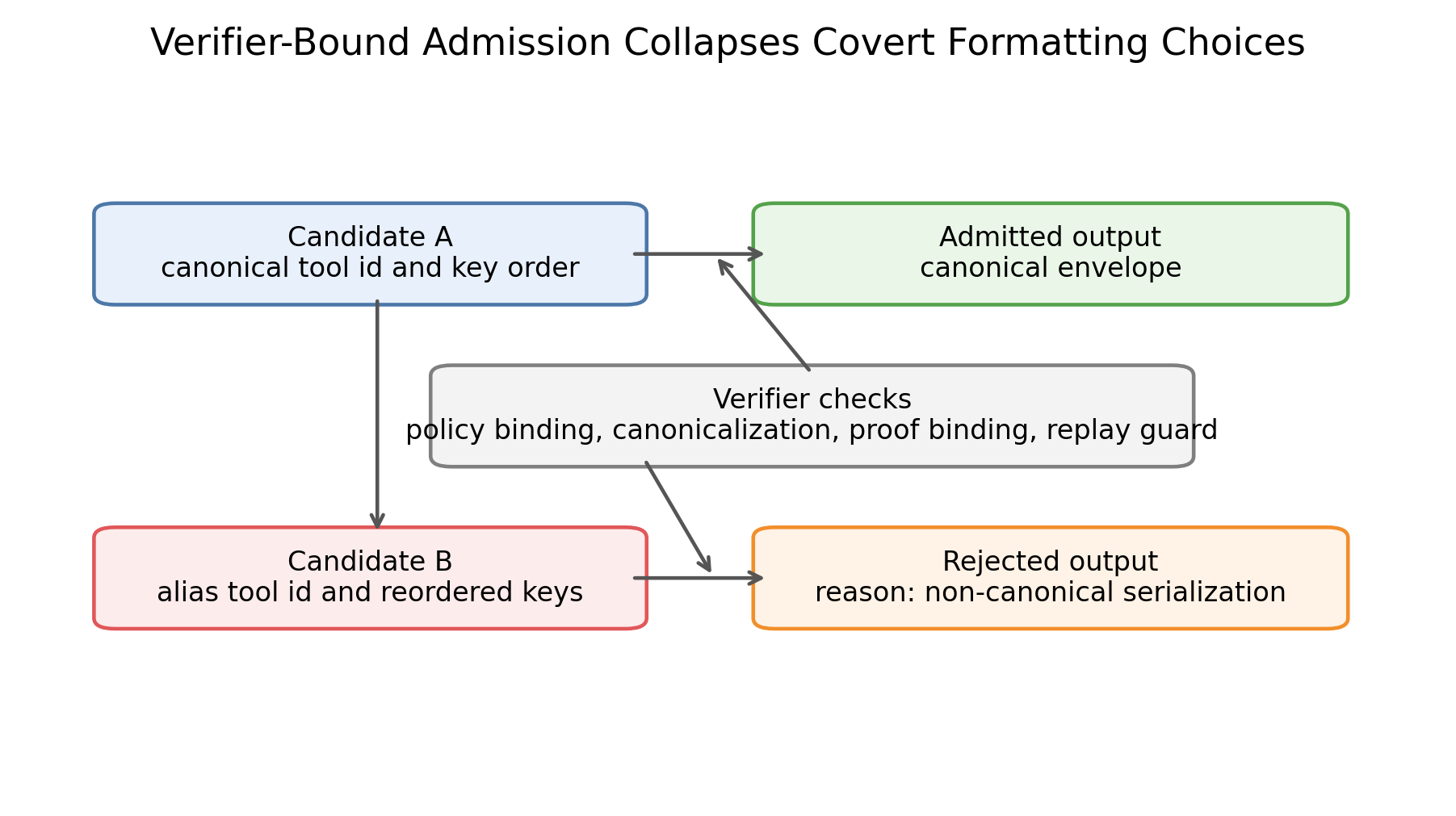}
\caption{Case study from our figure assets: covertly choosable variants collapse to canonical admitted form.}
\label{fig:case-study-appendix}
\end{center}
\end{figure}

A representative case-study output is shown in Figure~\ref{fig:case-study-appendix}. The figure contrasts an overt policy-compliant message, a covertly modified variant, and the admitted canonical form under verifier checks.

The core observation is structural: covert signaling opportunities often appear in ``small'' choices (field order, alias choice, optional metadata, tool-call normalization), not only in overt text content.

For a queue-status objective that requires one action and one confidence bucket, an overt candidate can be serialized as
\begin{verbatim}
{
  "action": "report",
  "priority": "high",
  "status": "revise"
}
\end{verbatim}
while a covert variant with the same surface intent can encode bits through aliasing and key order:
\begin{verbatim}
{
  "status": "revise",
  "priority": "high",
  "action": "notify"
}
\end{verbatim}
Under canonicalization and allowlist checks, the admitted form collapses to
\begin{verbatim}
{
  "action": "report",
  "priority": "high",
  "status": "revise"
}
\end{verbatim}
so alias and ordering freedom are removed from the admitted channel.

The same pattern appears in tool-heavy tasks. When a candidate uses semantically equivalent argument aliases or optional padding fields, fixed schema and canonical-argument checks reject the covert variant. This makes latent signaling through argument layout substantially harder without sacrificing deterministic execution semantics.

Case analysis also clarifies post-processing risks. If a span-replacement model rewrites admitted text and the rewrite is reintroduced without updated bindings, chain and proof checks reject the mutation on the next step. Timing and environment examples show similar behavior: fine-grained delay signaling collapses under bucket constraints, and noncanonical action permutations fail ordering and window checks. These traces are not substitutes for aggregate statistics, but they make failure modes concrete and help reviewers verify that specific predicate clauses remove specific channels.

\subsection{Measuring Perplexity: Oracle Language Model}\label{sec:measuring-ppl}
We report utility-oriented metrics rather than relying exclusively on perplexity, because covert signaling can occur while perplexity remains benign. Nevertheless, an oracle language model can provide a useful secondary lens.

Let $q_{\mathrm{oracle}}$ be a fixed evaluator model over admitted messages. For message sequence $m_{1:n}$,
\[
\mathrm{PPL}(m_{1:n}) = \exp\left(-\frac{1}{n}\sum_{i=1}^{n}\log q_{\mathrm{oracle}}(m_i\mid m_{<i})\right).
\]
In our pipeline, perplexity is interpreted as quality context, not as a covert-channel guarantee, and it is never used as an admission criterion. A channel can preserve perplexity while leaking through semantic choice, metadata, tool fields, or timing.

For this reason perplexity is paired with exact task success, slot-match, policy-violation rates, and decoder metrics. Structured and action-oriented families often require field-level correctness or execution outcomes that perplexity alone cannot capture.

If perplexity is reported, the oracle model and prompt wrapper are pinned and versioned; oracle changes are treated as benchmark changes with explicit reporting. Trend interpretation also joins perplexity with conformance logs and structural metrics to separate stylistic compression from genuine semantic failure.

\section{Detailed Threat Model}
We model a colluding sender-receiver pair interacting through an LLM-agent runtime with verifier boundary. The sender can adapt strategy over transcript history, choose among task-valid alternatives, and exploit any unchecked channel in envelope fields or runtime behavior.

Adversaries are assumed to have access to the same model outputs and tool-routing options as honest components, to adapt over long transcript horizons, and to optimize transform strategies through search-and-train loops. They may exploit message content, metadata, timing behavior, tool arguments, and environment interactions whenever those channels are insufficiently constrained. This is a high-adaptivity, high-collusion setting and is intentionally stronger than independent-attacker assumptions common in standard safety evaluations.

On the defender side, the protocol requires no-bypass verifier routing, correct canonicalization and proof checks, pinned policy and transcript-chain commitments, and explicit residual accounting for unresolved channels. Compromise of verifier roots of trust, physical side channels outside the protocol boundary, and universal guarantees over unseen workloads remain out of scope. Theorem-backed claims therefore apply to admitted transcripts within this boundary, while empirical claims remain scoped to evaluated model, task, and attacker classes.

Capability analysis is organized along knowledge, control, adaptivity, collusion, and persistence axes. The strongest evaluated profile is high in adaptivity/collusion and long-horizon persistence, with attacks grouped by mechanism (syntactic, tool-channel, timing, semantic-branch, and post-processing) and stage (pre-proof, in-proof, post-proof, post-admission).

The trust model is asymmetric: generator and post-processing components are untrusted, while verifier correctness and policy commitments are in boundary. The objective is bounded, auditable leakage rather than universal zero leakage; boundary failures or threshold violations invalidate affected security conclusions and require reruns with explicit incident scope and mitigation reporting.

\section{Detection Accuracy of Multinomial Sampling}
Although \CLBC is protocol-level, sampling dynamics still matter in empirical detection. For each seed and task family we estimate decoder advantage and MI proxies, pair them with utility signals, and compare per-slice and aggregate summaries against thresholds.

Strict lanes cluster near zero recoverability with narrow variance, while empirical lanes show wider variance and higher maxima, especially in high-choice families. This is consistent with semantic lower-bound pressure.

To avoid hiding weak slices, acceptance requires per-family and per-seed guards, minimum sample support, CI-width checks, and negative-control checks. Metrics are reported with margins, not pass/fail labels alone.

Decoder and MI estimates remain evidence proxies: decoder strength and feature choice can shift apparent recoverability, and MI uses finite-support approximations. We reduce instability via deterministic splits, bounded smoothing, and bootstrap checks for selected aggregates.

Between staged reruns, lightweight monitoring tracks reason-code frequencies, CI inflation, and baseline-gap erosion. Rapidly changing workloads also use temporal holdouts to detect silent drift in leakage or utility behavior.

We also report slice-level instability flags when rank ordering of baseline classes changes across adjacent windows without accompanying protocol changes. These reversals often indicate distribution shift, dataset composition drift, or decoder overfitting rather than genuine security improvement. Treating instability as a first-class signal improves triage quality and prevents premature claim upgrades.

\section{Minor Variations}
Several protocol variations preserve the core argument while changing engineering tradeoffs.

A common variant keeps mandatory per-turn verification while proving every $k$-th turn (or a policy-triggered subset), which often lowers median latency while retaining replay continuity and challenge verifiability. Another variant tightens latent intent and argument vocabularies for high-choice families, reducing semantic slack at potential utility cost. In practice these controls are most effective when paired with stronger tie-break enforcement, so that estimated branch counts above budget trigger either deterministic disambiguation or rejection.

Environment-aware variants move selected external actions into canonical transcript fields with allowlist, ordering, and batch-window checks. This shifts part of $L_t^{\mathrm{env}}$ from residual assumptions into online enforcement. Similar logic applies to timing channels: narrower policy windows can reduce residual timing bandwidth, although responsiveness may decline under bursty workloads.

Proof-system variants can combine lightweight attestations on all turns with periodic stronger receipts on audit windows, preserving accountability while improving throughput. Some deployments also expose challenger-triggered proof upgrades for selected transcript spans, increasing external trust without forcing strict proving for all traffic.

Operationally, these options are best managed through explicit assurance tiers that pair latency targets with required controls. Interpretation should track observed margins, not configuration labels: higher-assurance interpretations require baseline and attacker gaps to exceed predefined buffers. Because controls interact, evaluations should test coherent variation bundles rather than isolated toggles.

\section{Proof of Theorem \ref{maintheorem}}
Let $\mathcal{C}_t=(A_{1:T},R_{1:T},E_{1:t-1})$. By chain rule,
\[
\MI(S;\mathsf{T}_{1:T}\mid A_{1:T},R_{1:T})
=
\sum_{t=1}^{T}\MI(S;E_t\mid \mathcal{C}_t).
\]
From deterministic render up to residual,
\[
E_t = g_t(A_t,R_t,Z_t,U_t).
\]
By data processing,
\[
\MI(S;E_t\mid\mathcal{C}_t)
\le
\MI(S;Z_t,U_t\mid\mathcal{C}_t).
\]
Expanding the right-hand side gives
\[
\MI(S;Z_t,U_t\mid\mathcal{C}_t)
=
\MI(S;Z_t\mid\mathcal{C}_t)
+
\MI(S;U_t\mid\mathcal{C}_t,Z_t).
\]
By residual-budget validity,
\[
\MI(S;U_t\mid\mathcal{C}_t,Z_t)\le\mathcal{R}_t,
\]
thus
\[
\MI(S;E_t\mid\mathcal{C}_t)
\le
\MI(S;Z_t\mid\mathcal{C}_t)+\mathcal{R}_t.
\]
Summing over $t$ gives
\[
\MI(S;\mathsf{T}_{1:T}\mid A_{1:T},R_{1:T})
\le
\sum_{t=1}^{T}\MI(S;Z_t\mid\mathcal{C}_t) + \sum_{t=1}^{T}\mathcal{R}_t.
\]
This is exactly the theorem statement since $\mathcal{C}_t=(A_{1:T},R_{1:T},E_{1:t-1})$. If one absorbs public randomness into allowed information by setting \(A'_t=(A_t,R_t)\), the same expression can be written with \(A'_{1:T}\) notation.

The proof clarifies what must be true in implementation for theorem claims to be meaningful. Specifically, no-bypass admission and residual-budget validity are operational conditions that should be witnessed by verifier logs, conformance vectors, and boundary tests.

\begin{corollary}[Zero-residual special case]
If all residual channels are fully closed online (\(\mathcal{R}_t=0\) for all \(t\)), with
\(\mathcal{C}_t=(A_{1:T},R_{1:T},E_{1:t-1})\), then
\[
\MI(S;\mathsf{T}_{1:T}\mid A_{1:T},R_{1:T})
\le
\sum_{t=1}^{T}\MI(S;Z_t\mid \mathcal{C}_t).
\]
\end{corollary}

This corollary describes the target regime of complete online closure, where transcript leakage is controlled entirely by the latent-channel terms.

In practice the bound loosens when residual channels are only partially closed, policy commitments drift from deployed canonicalization, or seed handling introduces hidden reroll freedom. We therefore maintain channel-family residual declarations (timing, tool, tokenization, environment) and conservative upper bounds when overlap is uncertain.

Define raw candidate sequence $\hat{E}_{1:T}$ and admission operator $\Delta$ with $\mathsf{T}_{1:T}=\Delta(\hat{E}_{1:T})$. Under no-bypass routing, $\Delta$ is per-turn filtering by $\PiPred$, emphasizing that theorem claims are about admitted transcripts rather than raw generator proposals.

One subtle but important point is the conditioning set in each turn. The bound conditions on full allowed information and public randomness while carrying forward admitted history. This is what makes the statement robust to adaptive colluders: any strategy that reacts to past accepted messages is already represented in the conditional terms. Dropping history conditioning would produce cleaner notation but a weaker and potentially misleading guarantee for multi-turn settings.

Another practical implication is that residual terms should be interpreted as \emph{obligations}, not just bookkeeping. If a deployment chooses to leave a channel partially unresolved online, the resulting residual budget must be explicitly carried into claim language and decision thresholds. This is where the theorem interacts with governance: mathematical soundness alone does not prevent overclaiming unless residual declarations are enforced in reporting.

The proof can also be interpreted as a turn-wise accounting identity with conservative relaxation. The chain-rule decomposition isolates each admitted turn, data processing moves from observed envelope to latent-plus-residual variables, and residual-budget validity closes the inequality. Each step corresponds to a verifier-facing design decision, which is why implementation audits can map directly onto proof assumptions rather than relying on informal argument.

A useful diagnostic is to track empirical surrogates for the two summands separately across ablations. When latent terms shrink but residual proxies remain high, effort should prioritize boundary closure (timing/tool/environment). When residual proxies are stable but latent recoverability remains elevated, effort should prioritize intent schema and tie-break semantics. This decomposition is not only theoretically clean; it provides concrete engineering guidance for remediation.

From a verification perspective, this proof structure also motivates clause-specific regression tests. Each inequality step can be paired with a test family that intentionally violates exactly one corresponding assumption (for example, chain mismatch, seed mismatch, or residual-channel perturbation) and checks for deterministic rejection. Maintaining this mapping over code changes helps preserve theorem-to-implementation alignment as the system evolves.

\section{Proof of Proposition \ref{perpbound}}
Let \(\widetilde{\mathcal{C}}_t=(A_{1:T},R_{1:T},E_{1:t-1})\).
Define
\[
\begin{aligned}
\mathcal{L}_{Z,\min}(U)
=\inf_{\mathcal{P}:\mathcal{U}(\mathcal{P})\ge U}
\sum_{t=1}^{T}\MI_{\mathcal{P}}(S;Z_t\mid \widetilde{\mathcal{C}}_t).
\end{aligned}
\]
By definition of infimum, for any protocol $\mathcal{P}$ with utility at least $U$,
\[
\mathcal{L}_{Z,\min}(U)
\le
\sum_{t=1}^{T}\MI_{\mathcal{P}}(S;Z_t\mid \widetilde{\mathcal{C}}_t).
\]
Take $\mathcal{P}=\mathcal{P}_{\mathrm{CLBC}}$ and apply Theorem~\ref{maintheorem}:
\[
\begin{aligned}
\MI_{\mathcal{P}_{\mathrm{CLBC}}}(S;\mathsf{T}_{1:T}\mid A_{1:T},R_{1:T})
&\le
\sum_{t=1}^{T}\MI(S;Z_t\mid \widetilde{\mathcal{C}}_t) \\
&\quad + \sum_t\mathcal{R}_t.
\end{aligned}
\]
Set
\[
\begin{aligned}
\varepsilon_{\mathrm{model}}
&=
\sum_{t=1}^{T}\MI(S;Z_t\mid \widetilde{\mathcal{C}}_t)
-\mathcal{L}_{Z,\min}(U),\\
\varepsilon_{\mathrm{res}}
&= \sum_t\mathcal{R}_t.
\end{aligned}
\]
Substitution yields
\[
\MI_{\mathcal{P}_{\mathrm{CLBC}}}(S;\mathsf{T}_{1:T}\mid A_{1:T},R_{1:T})
\le
\mathcal{L}_{Z,\min}(U)+\varepsilon_{\mathrm{model}}+\varepsilon_{\mathrm{res}}.
\]
This proves the proposition.

The frontier proposition is intentionally non-constructive: it does not provide an exact closed-form $\mathcal{L}_{Z,\min}(U)$ for realistic workloads. Its value is interpretive, separating latent-policy suboptimality from residual-channel burden when comparing protocol configurations.

In practice we approximate the frontier by evaluating configurations with different strictness levels (e.g., schema-only, sampled proving, strict proving). Weaker admission semantics typically occupy higher-leakage regions at comparable utility, but these comparisons remain scoped to measured families, seeds, and model lanes.

Frontier plots should therefore include workload scope, seed count, model lane, and threshold margins. Policy or tool-schema changes that alter semantic branch structure require recomputing baselines before making cross-version comparisons.

The decomposition into $\varepsilon_{\mathrm{model}}$ and $\varepsilon_{\mathrm{res}}$ is operationally useful because it separates optimization debt from boundary debt. Large $\varepsilon_{\mathrm{model}}$ suggests better latent policies or render mappings may move the operating point, while large $\varepsilon_{\mathrm{res}}$ indicates unresolved channels that require protocol hardening rather than model tuning.

In empirical studies, this distinction helps avoid misdirected interventions. Teams often attempt to recover both utility and security by retraining alone, even when the dominant term is residual leakage from tool/timing/environment channels. The frontier proposition clarifies that these terms enter differently, so mitigation plans should target whichever term dominates in measured evidence.

For governance decisions, we recommend reporting not only point location on the frontier sketch but also movement direction after each mitigation cycle. Downward movement at stable utility supports stronger claim scope; horizontal movement with growing uncertainty may indicate overfitting to one workload slice; upward movement should trigger rollback or narrowed deployment scope.

\section{Impact of Watermarking on Model Factuality}\label{sec:factuality}
Factuality degradation is a central failure mode for constrained-generation systems. We evaluate it with task-grounded metrics (exact success, utility, and structured slot-match) rather than fluency-only proxies.

Constrained admission does not produce uniform factual collapse. Decision and summarization families remain relatively strong, while planning and tool-use families degrade more due to larger action/argument choice spaces. This points to semantic-choice pressure, not constraint presence alone, as the dominant driver.

Across families, stronger semantic schema design usually improves both factuality and leakage control by reducing ambiguous alternatives. By contrast, purely syntactic tightening can reduce quality without closing semantic channels. Effective mitigation targets choice structure, tie-break semantics, and proof-bound consistency jointly.

We track four recurring factuality failure classes: under-specification, over-canonicalization, tool-argument truncation, and tie-break mismatch. Because remedies differ by class, reports include per-family taxonomy counts, minimum sample support, and horizon-stratified summaries to surface long-transcript errors.

Mitigation follows a closed loop: schema refinement proposals, targeted conformance tests, controlled ablations, adaptive-attacker regression checks, and baseline reruns. Promotion requires leakage margins below thresholds, factuality floors above family-specific gates, no unresolved high-severity conformance failures, and archived reproducibility artifacts.

Factuality claims are scoped to evaluated families, models, seeds, and metric definitions. Post-deployment regressions trigger the same fail-closed update process used for leakage incidents: scope affected claims, rerun targeted evaluations, and publish updated margins before restoring broader conclusions.

To make these statements auditable, each family uses executable correctness contracts that specify required fields, admissible value sets, and transition constraints. We compute exact-match and partial-credit variants so that near-miss behavior can be analyzed without masking hard failures. This contract-style framing also simplifies regression triage: failures are localized to violated contract clauses rather than diffuse qualitative judgments.

We additionally run counterfactual stability checks in high-pressure families. Holding task context fixed, we perturb latent tie conditions and measure factual outcome variation. Large swings under small perturbations indicate unstable schema design or ambiguous tie-break policy; both conditions are undesirable because they degrade factual reliability and increase exploitable branch structure for covert signaling.

Finally, factuality review is integrated with security review rather than treated as a separate post hoc track. A mitigation that improves factuality but increases leakage margins is not automatically promotable, and vice versa. Promotion decisions therefore require joint satisfaction of factuality floors, leakage thresholds, and conformance health, with explicit residual declarations when unresolved channels remain.

\label{lastpagetotal}
\end{document}